\documentclass[numberedappendix]{emulateapj}
\usepackage{times}
\usepackage[normalem]{ulem}
\usepackage{graphicx}
\usepackage{natbib}
\usepackage{amsfonts}
\usepackage{amsmath}
\usepackage{multirow}
\usepackage{enumerate}
\usepackage{comment}
\usepackage[usenames]{xcolor}
\usepackage[plainpages=false, colorlinks=true, anchorcolor=blue, linkcolor=blue, citecolor=blue, bookmarks=false]{hyperref}
\newcommand{\be}{\begin{eqnarray}}
\newcommand{\ee}{\end{eqnarray}}

\newcommand{\lp}{\left(}
\newcommand{\rp}{\right)}

\newcommand{\slugcom}{Accepted for publication in The Astrophysical Journal}
\slugcomment{\slugcom}

\begin{document}

\title{Evidence for Cocoon Emission from the Early Light Curve of SSS17a}

\author{Anthony L. Piro and Juna A. Kollmeier}

\affil{The Observatories of the Carnegie Institution for Science, 813 Santa Barbara St., Pasadena, CA 91101, USA; piro@carnegiescience.edu}

\begin{abstract}
Swope Supernova Survey 2017a (SSS17a) was discovered as the first optical counterpart to the gravitational wave event GW170817. Although its light curve on the timescale of weeks roughly matches the expected luminosity and red color of an r-process powered transient, the explanation for the blue emission from high velocity material over the first few days is not as clear. Here we show that the power-law evolution of the luminosity, temperature, and photospheric radius during these early times can be explained by cooling of shock heated material around the neutron star merger. This heating is likely from the interaction of the gamma-ray burst jet with merger debris, so-called cocoon emission. We summarize the properties of this emission and provide formulae that can be used to study future detections of shock cooling from merging neutron stars. This argues that optical transient surveys should search for such early, blue light if they wish to find off-axis gamma-ray bursts and double neutron star gravitational wave events as soon as possible after the merger.
\end{abstract}


\section{Introduction}

On August 17, 2017, Advanced LIGO and Virgo detected gravitational waves from a binary neutron star merger for the first time in GW170817 \citep{GCN21509,GCN21513}. The Fermi satellite detected a coincident short gamma-ray burst, GRB170817A, and 10.87 hours later an optical counterpart was discovered, Swope Supernova Survey 2017a (SSS17a), in NGC 4993 \citep{GCN21529,Coulter2017}. The spectral energy distribution of SSS17a was observed within the next hour, from $g-$\citep{GCN21551} through $K-$bands with the Magellan telescopes. The first spectra of the source was also obtained within one hour and was then followed by almost daily optical spectra for the next week \citep{Shappee2017}. \citet{Drout17} presents bolometric light curves as well as individual bands over almost three weeks.

The spectra and light curves of SSS17a past the first few days roughly match what is expected for an \mbox{r-process} powered transient \citep{Shappee2017,Kilpatrick17}, also known as a macronova or kilonova \citep{Li1998,Kulkarni2005,Metzger2010,Roberts2011,Piran2013,Metzger2017}. However, at early times these observations revealed a nearly featureless blue component \citep{Shappee2017}. Two-component ejecta are not unexpected for neutron star mergers \citep{Perego2014,Fernandez2016}. A lanthanide-rich red component would come from material ejected on dynamical timescales via processes such as tidal forces \citep[e.g.,][]{Hotokezaka2013}. A lanthanide-free blue component could come out on longer timescales ($\sim$seconds) from accretion disk winds \citep{Fernandez2015}, {or from dynamically ejected shocked material at the interface between between merging neutron stars \citep[e.g.,][]{Bauswein13}.} The exact contribution of each component depends on the mass ratio of the merging binary, as well as the orientation relative to the line of sight \citep{Kasen2015}. Nevertheless, there are some difficulties with this interpretation of the blue component that we further discuss below. Chief among these are that the early ($\lesssim$3.5 days) bolometric luminosity does not appear to follow r-process heating, the inferred photospheric velocity is generally much higher than expected from disk winds, { and the sheer amount of dynamically ejected lanthanide-free material required strains this picture.}

Motivated by these issues,  we consider an alternative scenario for this event.  Specifically, for the early blue component, we advocate an origin in the cooling of shock-heated ejecta surrounding the neutron star merger. Such heating can occur due to the cocoon of material surrounding a gamma-ray burst \citep{Nakar17,Lazzati17,Gottlieb18}, which is crucial for collimating the jet \citep{Nagakura14,MurguiaBerthier14,MurguiaBerthier17}. Although there are some uncertainties in how the energy is distributed for this heating, we show that this can naturally explain the luminosity, temperature, and radius evolution of the early blue component.

In Section \ref{sec:motivation}, we discuss the observations for SSS17a, and { present simple arguments to motivate shock cooling as the origin for the early emission.} In Section~\ref{sec:model}, we present a model for cocoon emission and compare it to observations of SSS17a. In Section~\ref{sec:implications}, we discuss implications for electromagnetic counterparts to gravitational waves the surveys searching for them. We then conclude in Section~\ref{sec:conclusions} with a summary of our work and a discussion of future observations.

\section{Motivating Arguments}
\label{sec:motivation}

We begin by highlighting the observed optical/infrared  properties of SSS17a that drive our thinking about the physical nature of this event. This is followed by simple energetics and scaling arguments that suggest a shock cooling model is a possible solution.

\subsection{Optical/Infrared Properties of SSS17a}
\label{sec:properties}

\begin{figure}
\epsscale{1.2}
\plotone{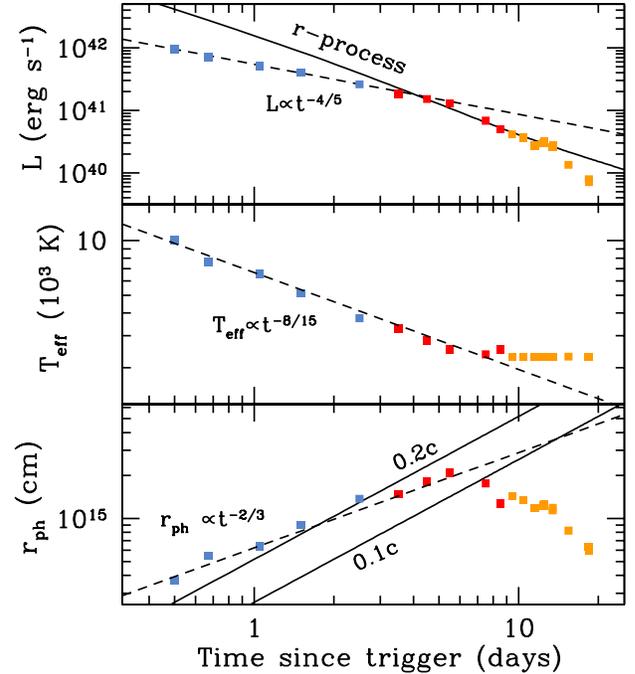}
\caption{The bolometric luminosity, effective temperature, and photospheric radius of SSS17a \citep{Drout17}. Epochs where multi-band photometry is used to derive these properties are plotted in blue or red before and after $\sim$3.5 days, respectively. Orange points only have $K-$ or $H-$bands available and assume an effective temperature of $2300\,{\rm K}$. Solid lines show r-process heating for $0.05\,M_\odot$ of material with the thermalization efficiency of \citet{Barnes2016}. Dashed lines highlight the power-law evolution of the transient during the first $\sim$ 3.5 days, which is consistent with $s=-d\ln E/d\ln v\approx3$ shock heated ejecta.}
\label{fig:ltr}
\end{figure}

The luminosity, effective temperature, and photospheric radius of SSS17a from \citet{Drout17} are summarized in Figure \ref{fig:ltr}.  There are several features of importance in this data series.  After $\sim$3.5 days, shown by red and yellow points in Figure \ref{fig:ltr}, the emission is well-described by heating due to r-process material, shown as the solid line in the figure.  To generate this line, we assumed  $0.05\,M_\odot$ of r-process material and the characteristic $t^{-1.3}$ power-law heating \citep{Metzger2010,Roberts2011}, { including small corrections due to the changing thermalization \citep[using the parameterization provided in][we set $a=0.27$, $b=0.10$, and $d=0.60$]{Barnes2016}.  The thermalization is roughly constant at $\approx$25\% at late times.} This is also consistent with the optical and infrared spectra, which resemble standard kilonova models \citep{Shappee2017,Kilpatrick17,Chornock17,Kasliwal17,Smartt17}.  From this data, it is thus clear that r-process heating can explain the late optical/infrared data.

The emission before $\sim$3.5 days (blue points in Figure~\ref{fig:ltr}) is very blue and thus must be lanthanide-free to correctly match the color.  Such material may be from accretion disk wind \citep{Metzger2014,Kasen2015} or dynamically ejected \citep{Wanajo2014,Shappee2017,Kilpatrick17,Kasen2017}.  From a snapshot of the color alone, one could potentially conclude that this material represents r-process heating as well.  The time evolution during these times, however, appears to severely constrain this picture.  First, the early light curve is actually considerably shallower than expected from r-process heating. { This is expected naturally around peak for any radioactively powered transient. However, the shallow portion should last of order the time of the rise (as this is set by the diffusion timescale, see, for example, the light curves in \citealp{Roberts2011}). In the case of SSS17a though the rise was less than 11 hrs and the shallow decay lasted for days.} Second, the disk-wind model in particular has the problem that the wind velocities are generally much lower ($\lesssim0.1c$, \citealp{Fernandez2015}) than what we infer in the bottom panel of \mbox{Figure \ref{fig:ltr}}. { Third, dynamically ejected material shocked at the interface between the neutron stars can potentially be lanthanide-free and high velocity, but the mass of this material is generally $\lesssim0.01\,M_\odot$ \citep{Sekiguchi16}. The high early luminosity of SSS17a would require a factor of $\sim2-3$ more mass than this.} Finally, the power-law behavior for $L$, $T_{\rm eff}$, and $r_{\rm ph}$ over a similar timescale begs explanation.

\subsection{Cooling of Shock-Heated Material}
\label{sec:shocks}

{ We next discuss the energetics needed if the early emission is from the cooling of shock-heated material. Consider material with mass $M$ and radius $R$, which is heated by a shock that deposits an energy $E$.} This characteristic luminosity as this material cools is expected to scale as
\be
    L \approx \frac{4\pi cR}{\kappa} \frac{E_{\rm sh}}{M},
    \label{eq:lum}
\ee
where $\kappa$ is the specific opacity (see Appendix \ref{sec:derivation} for a derivation of this expression). This scaling proportional to radius is a generic prediction of shock cooling \citep{Nakar10,Piro13,Nakar14}. This is because a larger initial radius results in less adiabatic cooling during expansion. { The typical scale of ejecta when the jet breaks out is $\sim10^{10}\,{\rm cm}$, similar to the physical scale derived from the gamma-ray emission \citep{Nagakura14} and roughly matches a jet traveling at $\sim\,c$ during the 1.7~sec between the merger and the GRB. From this we can then estimate the energy of the shock for an assumed mass, opacity and luminosity
\be
    E_{\rm sh} \approx 5\times10^{50}\kappa_{0.1}L_{42}M_{0.01}R_{10}^{-1}\,
    {\rm erg},
\ee
where $\kappa_{0.1}=\kappa/0.1\,{\rm cm^2\,g^{-1}}$, $L_{42}=L/10^{42}\,{\rm erg\,s^{-1}}$, $M_{0.01}=M/0.01\,M_\odot$, and $R_{10}=R/10^{10}\,{\rm cm}$. Note that this low opacity means that relatively lanthanide-free material is still required to explain the color of this emission, but this is expected for the ejecta closest to the merger from processes such as disk winds and shocked material \citep{Bauswein13,Metzger2014,Perego2014}.}

In comparison, this same material has an associated kinetic energy of
\be
    E_{\rm KE} \approx Mv^2/2 \approx 4\times10^{50}M_{0.01}v_{0.2}^2\,{\rm erg},
\ee
where $v_{0.2}=v/0.2c$ is set by the photospheric velocity measurements in Figure \ref{fig:ltr}. Of course there could be factor of a few differences in the exact values of $R$, $\kappa$, and $v$. Nevertheless, the near equality of $E_{\rm sh}$ and $E_{\rm KE}$ argues that whatever is imparting the velocity is also heating the material. Heating sources like radioactivity would not naturally result in this similarity, but shock cooling could.

We next follow \citet{Nakar17} and consider a shocked envelope that has an energy distribution $dE/dv \propto v^{-s}$, assuming $s>-1$. If $m(>v)$ is the total mass with velocity above $v$, then this scalings implies $m(>v)\propto v^{-(s+1)}$. This is a convenient way to parameterize our ignorance about how the jet deposits its energy into the ejecta. \citet{Nakar17} quote that numerical simulations suggest roughly constant energy per logarithm of velocity, or $s=1$, but there is sufficient uncertainty that we consider a range of values for $s$ in our work.

Combining this energy distribution with the equations for adiabatic expansion and radiative cooling, we find
\be
    &&L\propto t^{-4/(s+2)},\label{eq:ls}
    \\
    &&T_{\rm eff} \propto t^{1/(s+3)-1/(s+2)-1/2},
    \\
    &&r_{\rm ph} \propto t^{(s+1)/(s+3)},
\ee
where we use an opacity that is temperature and density independent (an assumption that is further discussed below).
For the full derivation, see Section \ref{sec:model}. Thus, if the early light curve is determined by shock cooling, the luminosity, effective temperature, and photospheric radius must all be consistent with the same value of $s$. Comparing with the SSS17a, we find that $s=3$ is consistent with the data (which is plotted as dashed lines in Figure \ref{fig:ltr}). This is another strong argument for a shock cooling explanation.

\section{Cocoon Emission Model}
\label{sec:model}

Motivated by the good match to shock cooling, we derive the basic properties of a cocoon heating model. This closely follows the analytic framework discussed in \citet{Nakar17}, but with additional details to facilitate comparisons with the SSS17a data.

The basic picture is that before the GRB punches through the $\sim0.01\,M_\odot$ of debris around the merger remnant, this material has expanded out to a radius of $R\sim10^{10}\,{\rm cm}$. This imparts a self-similar energy distribution with velocity $dE/dv\propto v^{-s}$, so that the mass profile with velocity is
\be
    m(>v) \approx M \lp\frac{v}{v_0} \rp^{-(s+1)},
    \label{eq:mass}
\ee
where $M$ is the total shocked mass, and $v_0$ is the minimum velocity in the ejecta.

An observer sees a luminosity from the depth where the diffusion time is roughly equal to the time of the observation (the so-called ``diffusion depth'' or ``luminosity depth,'' \citealp{Nakar10}). This corresponds to an optical depth
\be
    \tau \approx \frac{\kappa m}{4\pi v^2 t^2} \approx \frac{c}{v}.
    \label{eq:tau}
\ee
Converting factors of $v$ to $m$ using Equation (\ref{eq:mass}), we solve for the diffusion depth
\be
    m_{\rm diff} = M\lp \frac{t}{t_{\rm diff}} \rp^{2(s+1)/(s+2)},
    \label{eq:mdiff}
\ee
where
\be
    t_{\rm diff} = \lp \frac{\kappa M}{4\pi c v_0}\rp^{1/2}
    = 0.5\,\kappa_{0.1}^{1/2}M_{0.01}^{1/2}v_{0.1}^{-1/2}\,{\rm days},
\ee
is the characteristic diffusion timescale with $v_{0.1}=v/0.1c$.

The initial energy at a given depth is $E_0 \approx mv^2/2$, but this then adiabatically cools as $1/r$, so that
\be
    E(t) \approx E_0 \lp \frac{R}{vt}\rp
    \approx \frac{1}{2} \frac{mvR}{t}
\ee
where $R$ is a characteristic initial radius. A single radius can be used here because the initial range of radii of the material is much smaller than the range at later times once the material has expanded. The luminosity is simply $L\approx E(t)/t$, thus substituting $m_{\rm diff}$ using Equation (\ref{eq:mdiff}) results in
\be
    L \approx \frac{Mv_0R}{2t_{\rm diff}^2}
    \lp \frac{t}{t_{\rm diff}} \rp^{-4/(s+2)},
    \label{eq:l}
\ee
for the time dependent luminosity.

To derive scalings at the photospheric depth, we set $\tau\approx1$. Using Equation (\ref{eq:tau}), we derive
\be
    v_{\rm ph} = v_0 \lp \frac{t}{t_{\tau=1}} \rp^{-2/(s+3)},
\ee
where
\be
    t_{\tau=1} = \lp\frac{c}{v_0} \rp^{1/2}t_{\rm diff}
    = 1.5\, \kappa_{0.1}^{1/2}M_{0.01}^{1/2}v_{0.1}^{-1}\,{\rm days},
\ee
is the characteristic time for the material to become optically thin. The photospheric radius is in turn
\be
    r_{\rm ph} = v_{\rm ph}t = v_0 t_{\tau=1} \lp\frac{t}{t_{\tau=1}} \rp^{(s+1)/(s+3)}.
    \label{eq:rph}
\ee
Finally, combining Equation (\ref{eq:l}) and (\ref{eq:rph}), we find
\be
    T_{\rm eff} &\approx& \lp \frac{L}{4\pi r_{\rm ph}^2\sigma_{\rm SB}} \rp^{1/4}
    \propto t^{1/(s+3)-1/(s+2)-1/2},
\ee
for the effective temperature evolution.

For $s=3$ we find the following scalings, for the luminosity, effective temperature, and photospheric radius,
\be
    &&L = 9.5\times10^{40}\kappa_{0.1}^{-3/5}M_{0.01}^{2/5}
        v_{0.1}^{8/5}
        R_{10}
        t_{\rm day}^{-4/5}\,{\rm erg\,s^{-1}},
        \\
    &&T_{\rm eff} = 6.2\times10^3\kappa_{0.1}^{-7/30}
    M_{0.01}^{1/60}
        v_{0.1}^{1/15}
        R_{10}^{1/4}
        t_{\rm day}^{-8/15}\,{\rm K},
    \\
    &&r_{\rm ph} = 3.0\times10^{14}
        \kappa_{0.1}^{1/6}M_{0.01}^{1/6}
        v_{0.1}^{2/3}
        t_{\rm day}^{2/3}\,{\rm cm}.
\ee
In Appendix \ref{sec:selfsimilar} we summarize the scalings for other values \mbox{of $s$} for comparison to future observations.

Unfortunately, the number of free parameters is sufficiently large that they cannot be all solved uniquely with just $L$, $T_{\rm eff}$, and $r_{\rm ph}$. For Figure \ref{fig:ltr} in particular, { we use $\kappa=0.5\,{\rm cm^2\,g^{-1}}$, $M=0.01\,M_\odot$, $v_0=0.2c$, and $R=5.0\times10^{10}\,{\rm cm}$, but a range of similar values would also roughly match the data.} Nevertheless, the fact that a good match to the early time data can be made with a physically reasonable set of values strengthens the claim that cocoon emission is determining the early luminosity.

{This particular set of parameters implies that the kinetic energy of this material is $\sim Mv_0^2/2=4\times10^{50}\,{\rm erg}$}. In comparison to the typical beaming-corrected short GRB energies \citep{Fong2015}, this is on the high side, but not unreasonably large. Plus it is not exactly known how much energy a GRB supplies to its surrounding medium versus the jet itself anyway.

In Appendix \ref{sec:nonisotropic}, we discuss how these results change if the shock heating and emission are confined to a half angle $\theta$. {The main conclusion is that $r_{\rm ph}$ remains unchanged and $T_{\rm eff}$ is only different by a little more than $10\%$. The isotropic equivalent luminosity (the luminosity that would be observed) is smaller than that given by Equation (\ref{eq:l}), but this can be accommodated by changing the parameters somewhat. The main difference is that $M$ now refers to the isotropic equivalent mass and the actual shocked mass is $\approx M\theta^2/2$. For a typical opening angle of $\theta\sim0.5$ \citep{Fong2015}, this would give a shocked mass of $\approx 1\times10^{-3}\,M_\odot$ and a shock energy of $\approx M\theta^2v_0^2/4\approx 4\times10^{49}\,{\rm erg}$. Both of these values are more in line with what one might expect for an average short GRB, perhaps arguing that the cocoon emission is not isotropic. A larger collection of future observations will help us better understand if there are viewing angle differences.}

\section{Implications for Counterpart Searches}
\label{sec:implications}

\begin{figure}
\epsscale{1.2}
\plotone{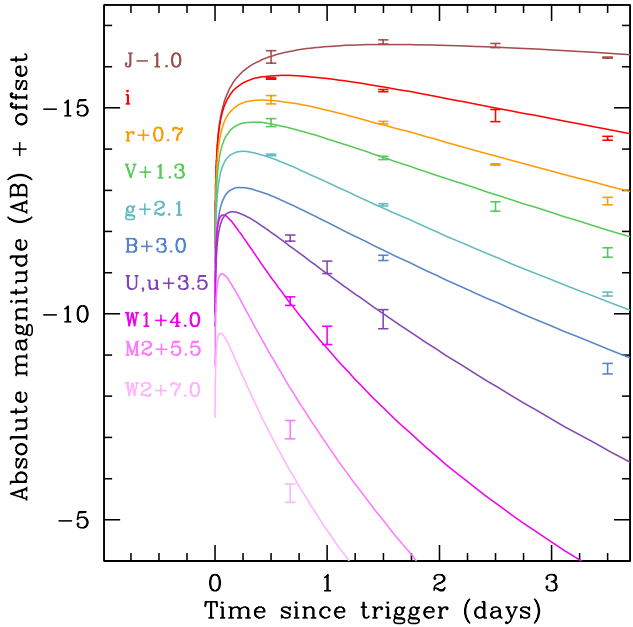}
\caption{{ Photometry from \citet{Drout17} and \citet{Shappee2017} in comparison to a cocoon model using the same parameters as in Figure \ref{fig:ltr}. This shows that the shock cooling would be expected to peak at $\sim3-6\,{\rm hrs}$ in the bluer bands.}}
\label{fig:photometry}
\end{figure}

While transients have been studied for decades, the combination of gravitational waves and rapid multi-wavelength electromagnetic follow-up provides an opportunity to probe the physics of these events in fundamentally different ways.  For future wide field transient surveys, an important goal will be catching the remnant as close as possible following merger. If SSS17a is typical, this work argues that cooling from shock heating is likely the first ultraviolet/optical signal that can be seen, at least when viewed off-axis.

SS17a had an isotropic equivalent luminosity of \mbox{$\approx5\times10^{41}\,{\rm erg\,s^{-1}}$} at one day after merger. For a survey like the Zwicky Transient Facility \citep[ZTF,][]{Law2009}, which could image down to a magnitude of $21$ over roughly $3\pi$ of the sky most nights, these events will be observable out to $\approx200\,{\rm Mpc}$. For a beam-corrected rate of short GRBs of $300\,{\rm Gpc^{-3}\,yr^{-1}}$ \citep{Fong2015}, this would give roughly $\approx7$ SSS17a-like events per year. This estimate could change drastically depending on the relative beaming of the shock emission and the relativistic jet itself, so that constraining this rate could have interesting implications for understanding the geometry of short GRBs. For the Large Synoptic Survey Telescope \citep{lsst}, going down to a magnitude of $24.5$ but with merely a 3 days cadence, could  detect many more of these events at roughly $\approx300 $ per year. But given how quickly these events evolve compared to a 3 day cadence, it is not clear how they would actually be identified for study.

Also, interesting is how hot SSS17a was at early times. { In Figure~\ref{fig:photometry}, we compare the shock cooling model from Figure~\ref{fig:ltr} to the photometry from \cite{Drout17} and \citet{Shappee2017}. Shock cooling predicts that in the bluer bands SSS17a would peak at $\sim3-6\,{\rm hrs}$, which could prove an important discriminant from radioactively powered models of the early emission. The $U-$band had an AB magnitude of $-16$ at peak, about $0.5$ magnitudes brighter than the $i-$band at peak, even though short-cadence transient surveys plan to look in redder bands motivated by the red colors expected, and now seen, for \mbox{r-process} powered transients.}  While atmospheric absorption would make a $U-$band transient survey prohibitive on the ground, these events might be especially attractive for space based missions. An ultraviolet survey sensitive to a magnitude of $21.5$ \citep{ultrasat} and imaging roughly half of the sky with a regular cadence could see these events out to $\approx320\,{\rm Mpc}$ with a rate of $\approx20$ per year. Especially important would be a short sub-day cadence, so that the fast evolution of the event can be studied.

An additional issue is the gamma-ray emission from this event and whether it truly is an off-axis short GRB. The cocoon emission we highlight here could be accompanied by shock breakout, which would also appear in gamma-rays \citep{Nakar2015,Nakar17,Bromberg18}. Since such a breakout would be mildly relativistic, the impact of pair-generation naturally gives a breakout temperature of $60\,{\rm keV}\lesssim T \lesssim 200\,{\rm keV}$ \citep{Katz2010}, similar to what was observed for the gamma-ray counterpart GRB170817A \citep{Goldstein2017}. If this is the case, then the gamma-ray emission would be more isotropic than how it falls off for an off-axis short GRB, which would increase the chances for gamma-ray searches to find these events. In the future, as more events are found, the gravitational wave measurements will inform us on the orientation of the binaries so that it can be better understood whether we are seeing off-axis ultra-relativistic jets or shock breakout from a cocoon.

\section{Summary and Conclusions}
\label{sec:conclusions}

{ We studied the early $\lesssim4\,{\rm day}$ optical/infrared emission from SSS17a, and argued that its luminosity is may be explained by shock cooling emission.} Motivated by this, we presented an analytic model inspired by the work of \citet{Nakar17} and \citet{Gottlieb18} on cocoon emission. We generalized it to consider a range of energy deposition profiles $dE/dv\propto v^{-s}$, and find that SSS17a's luminosity, temperature, and photospheric radius evolution best match $s=3$. We also provided solutions for a range of values for $s$ for comparison to future gravitational wave counterparts. { The shocked mass and energy $\approx0.01\,M_\odot$ and $\approx4\times10^{50}\,{\rm erg}$, respectively, needed to match the early emission are generally consistent with what one might expect for heating by a relativistic jet. We also discuss how these parameters change if the shock heating is non-isotropic. Although radioactively powered models may also explain this early blue emission (though note the potential difficulties summarized in Section \ref{sec:properties}), shock cooling differs in that it predicts the emission to be even bluer and brighter at earlier times. This will hopefully be tested with future observations of gravitational wave counterparts.}

Over the last decade, there has been increasing work on theoretically exploring the electromagnetic counterparts to neutron star mergers motivated by the anticipated detection of such events by Advanced LIGO and Virgo \citep[for example, see the summary by][]{Metzger12}.  With the detection of GW170817/SSS17a, these models can now, for the first time, be confronted with a rich observational dataset. Two of the main foci of these predictions have been counterparts that are (1) more isotropic than the highly beamed gamma-ray emission and (2) early on so that they can be seen rapidly post-merger. This work argues that the cocoon emission, as put forth by \citet{Nakar17} and \citet{Gottlieb18}, may be the theoretical counterpart that best predicted the early-time  emission. This means that searches in bluer colors rather than the infrared would be a better strategy for identifying off-axis short GRBs and gravitational wave counterparts in the shortest time post-merger. A space-based mission in the ultraviolet would be especially well-suited for finding these events.

In the future, with a larger sample of cocoon emission events, the environment of neutron star mergers and the interaction of the short GRB jet with this material can be better understood. The shock cooling emission is a strong probe of the radius and mass of the surrounding material, and this can be compared to the mass ratio of the binary measured from gravitational waves to look for correlations. The energy associated with emission and the kinetic energy supplied to the material allows us to measure how much of the relativistic jet's energy is deposited into its environment as it tunnels out. A higher value of $s$ indicates that more energy is deposited at the base of the jet. Using the formulae in Appendix \ref{sec:selfsimilar}, different values of $s$ can be inferred and compared with various properties of the binary and associated short GRB. 

While SSS17a is only a single event, the enthusiasm and dedication of the electromagnetic community in following up this source provides a glimpse (for better or worse) of what is possible when coordinated, detailed observations are performed on events like these and the key interplay between ground-based and space-based facilities.  This is hopeful news as time-domain astronomy enters its next phase of maturation, one in which we should expect theoretical models of these events to be directly tested.

\acknowledgments
ALP thanks Ehud Nakar for early discussions on shock-cooling. We additionally thank Ehud Nakar and Tsvi Piran for their visionary work on cocoon emission as well as the One-Meter Two-Hemisphere/Swope Supernovae Survey (1M2H/SSS) collaboration. 

\begin{appendix}

\section{One Zone Shock Cooling Model}
\label{sec:derivation}

Assuming that the shocked ejecta can be described as a single zone with mass $M$, velocity $v$, and radius $r=R+vt$, we can derive the simple scaling for shock cooling emission. The gamma-ray burst jet shocks an amount of material $M$.  If at a given time the ejecta has energy $E$, then the observed luminosity is
\be
    L = \frac{E}{t_{\rm diff}},
\ee
where
\be
    t_{\rm diff} = \tau\frac{r}{c},
    \label{eq:tdiff}
\ee
is the diffusion timescale and $\tau$ is the optical depth
\be
    \tau = \frac{\kappa M}{4\pi r^2}.
\ee
The luminosity at any given time is then
\be
    L = \frac{4\pi r c}{\kappa } \frac{E}{M}.
\ee
For material with adiabatic index of $\gamma=4/3$, the internal energy scales as $E\propto 1/r$. Therefore, we can scale the energy and radius to the initial energy of the shock and the initial radius, $E/E_{\rm sh}= R/r$. This results in
\be
    L = \frac{4\pi R c}{\kappa } \frac{E_{\rm sh}}{M},
\ee
for the shock cooling luminosity as used in Equation (\ref{eq:lum}). This is the same general scaling applicable to the early emission from supernovae \citep{Nakar10,Piro13,Nakar14}.

\section{Self-similar Solutions}
\label{sec:selfsimilar}

In Section \ref{sec:model}, we focused on the scalings for cocoon emission with $s=3$. In the future, hopefully additional neutron star binaries will be identified closely after merger so that the cocoon emission can be observed. In principle, how the jet deposits its energy into the ejecta can vary. Therefore here we present the scalings for different values of $s$.

For the case of $s=1$,
\be
    &&L = 6.5\times10^{40}\,\kappa_{0.1}^{-1/3}M_{0.01}^{2/3}
        v_{0.1}^{4/3}
        R_{10}
        t_{\rm day}^{-4/3}\,{\rm erg\,s^{-1}},
    \\
    &&T_{\rm eff} = 5.4\times10^3\,\kappa_{0.1}^{-5/24}
    M_{0.01}^{1/24}
        v_{0.1}^{1/12}
        R_{10}^{1/4}
        t_{\rm day}^{-7/12}\,{\rm K},
    \\
    &&r_{\rm ph} = 3.2\times10^{14}\,
        \kappa_{0.1}^{1/4}M_{0.01}^{1/4}
        v_{0.1}^{1/2}
        t_{\rm day}^{1/2}\,{\rm cm}.
\ee
For the case of $s=2$,
\be
    &&L = 8.2\times10^{40}\,\kappa_{0.1}^{-1/2}M_{0.01}^{1/2}
        v_{0.1}^{3/2}
        R_{10}
        t_{\rm day}^{-1}\,{\rm erg\,s^{-1}},
    \\
    &&T_{\rm eff} = 5.9\times10^3\,\kappa_{0.1}^{-9/40}
    M_{0.01}^{1/40}
        v_{0.1}^{3/40}
        R_{10}^{1/4}
        t_{\rm day}^{-11/20}\,{\rm K},
    \\
    &&r_{\rm ph} = 3.1\times10^{14}\,
        \kappa_{0.1}^{1/5}M_{0.01}^{1/5}
        v_{0.1}^{3/5}
        t_{\rm day}^{3/5}\,{\rm cm}.
\ee
Finally, for the case of $s=4$,
\be
    &&L = 1.0\times10^{41}\,\kappa_{0.1}^{-2/3}M_{0.01}^{1/3}
        v_{0.1}^{5/3}
        R_{10}
        t_{\rm day}^{-2/3}\,{\rm erg\,s^{-1}},
    \\
    &&T_{\rm eff} = 6.4\times10^3\,\kappa_{0.1}^{-5/21}
    M_{0.01}^{1/84}
        v_{0.1}^{5/84}
        R_{10}^{1/4}
        t_{\rm day}^{-11/21}\,{\rm K},
    \\
    &&r_{\rm ph} = 2.9\times10^{14}\,
        \kappa_{0.1}^{1/7}M_{0.01}^{1/7}
        v_{0.1}^{5/7}
        t_{\rm day}^{5/7}\,{\rm cm}.
\ee
These will hopefully be useful for comparisons to future observations.

\section{Non-isotropic Heating and Emission}
\label{sec:nonisotropic}

The main derivation of cooling in Section \ref{sec:model} assumes the shocked material is roughly isotropic. Here we discuss what corrections non-isotropic heating may have.

Consider that the GRB jet only shocks material with half angle $\theta$ (note that this could in principle be much wider than the GRB jet), so that a mass $M\theta^2/2$ is shocked (where $M$ now refers to the isotropic equivalent mass). Then the mass profile with velocity is
\be
    m(>v) \approx \frac{M\theta^2}{2} \lp\frac{v}{v_0} \rp^{-(s+1)}.
    \label{eq:mass2}
\ee
The optical depth is
\be
    \tau \approx \frac{\kappa m}{2\pi \theta^2v^2t^2}
   \approx \frac{\kappa M}{4\pi v_0^2t^2} \lp\frac{v}{v_0} \rp^{-(s+3)},
    \label{eq:tau2}
\ee
so in fact the optical depth is not changed.
Converting factors of $v$ to $m$ using Equation (\ref{eq:mass2}), we solve for the diffusion depth
\be
    m_{\rm diff} = \frac{M\theta^2}{2}\lp \frac{t}{t_{\rm diff}} \rp^{2(s+1)/(s+2)},
\ee
where again $t_{\rm diff}$ is unchanged from before.

{ A key difference in the non-isotropic case is that the shocked material can spread laterally as it expands from the merger site. The initial shocked material is a cone with volume $\approx \pi (R\theta)^2R/3$, which then expands into a hemisphere with volume $2\pi (vt)^3/3$. The energy at any given time then scales as the cube root of this volume ratio, thus
\be
    E(t) \approx E_0 \lp \frac{R}{vt} \rp \lp\frac{\theta^2}{2} \rp^{1/3}.
\ee
Following the discussion in Section \ref{sec:model}, the resulting luminosity is
\be
    L \approx \lp \frac{\theta^2}{2}\rp^{4/3}\frac{Mv_0R}{t_{\rm diff}^2}
    \lp \frac{t}{t_{\rm diff}} \rp^{-4/(s+2)}.
\ee
Although this luminosity is decreased by $(\theta^2/2)^{4/3}$ from the isotropic case, an observer would see the isotropic equivalent luminosity
\be
    L_{\rm iso} = 2L/\theta^2,
\ee
so that this is just different from Equation (\ref{eq:l}) by $(\theta^2/2)^{1/3}$. For a typical opening angle of $\theta\sim0.5$ \citep{Fong2015}, this is a $\sim50\%$ change in the isotropic luminosity. Given the presently large uncertainties in some of the parameters we fit, this difference can be accommodated with parameter changes.  A clear observational goal to test this picture will be to restrict these parameters.  Since the optical depth is not changed, neither is the photospheric radius. The effective temperature is changed by $(\theta^2/2)^{1/12}$, or $\sim16\%$ for $\theta\sim0.5$.}

The main conclusion is that non-isotropic emission does not make a large difference in the comparisons to SSS17a because we will always necessarily be comparing to the isotropic equivalent luminosity anyway. The main difference physically is that $M$ now refers to the isotropic equivalent mass, with the actual shock heated mass being $\approx M\theta^2/2$ with an associated shock energy $\approx M\theta^2v_0^2/4$. This may potentially help the model better match the energetics of short GRBs.
\vspace{0.5cm}

\end{appendix}

\bibliographystyle{apj}

\end{document}